\documentclass[prb,showpacs,preprintnumbers,amsfonts,amssymb,amsmath,floats,twocolumn,aps]{revtex4}

\usepackage{graphicx}
\usepackage{dcolumn}
\usepackage{bm}
\usepackage[final,dvips]{epsfig}
\usepackage{bm}
\usepackage{color}

\begin{document} 
  
\title[Self-energy-functional approach to systems of correlated electrons]
      {Self-energy-functional approach to systems of correlated electrons} 

\author{Michael Potthoff \cite{ad}}

\affiliation{
Lehrstuhl Festk\"orpertheorie,
Institut f\"ur Physik, 
Humboldt-Universit\"at zu Berlin, 
10115 Berlin, 
Germany
}
 
\begin{abstract}
The grand potential of a system of interacting electrons is considered 
as a stationary point of a self-energy functional.
It is shown that a rigorous evaluation of the functional is possible for 
self-energies that are representable within a certain reference system. 
The variational scheme allows to construct new non-perturbative and 
thermodynamically consistent approximations.
Numerical results illustrate the practicability of the method.
\end{abstract} 
 

\maketitle 

\section{Introduction}
\label{sec:intro}

Systems of strongly correlated electrons continue to represent a central 
subject of current research.
Different interesting correlation phenomena, 
such as high-temperature superconductivity, \cite{OM00} 
Mott metal-insulator transitions \cite{Geb97} 
or itinerant ferromagnetism, \cite{BDN01} 
are far from being finally clarified.
Progress in this field crucially depends on the development of new 
theoretical methods as even highly idealized model systems 
pose notoriously difficult problems.
There are only a few general approaches 
which are able to access the equilibrium thermodynamics as well as 
excitation properties of an extended system of correlated electrons.

General methods can be based on the Green's-function formalism of Luttinger 
and Ward \cite{LW60} and Baym and Kadanoff: \cite{BK61}
Here the grand potential $\Omega$ is expressed in terms of the time- or 
frequency-dependent one-electron Green's function ${\bf G}$.
The functional $\Omega[{\bf G}]$ can be shown to be stationary at the 
physical ${\bf G}$.
In principle, this is an exact variational approach which provides 
information not only on static equilibrium but also on dynamic excitation 
properties.
The functional dependence $\Omega[{\bf G}]$, however, is generally not known
explicitly as it must be constructed by summation of an infinite series of 
renormalized skeleton diagrams.
In the standard approximation the exact but unknown functional is replaced 
by an explicitly known but approximate one which is based on an incomplete 
summation of the diagram series.
This leads to the well-known perturbational (``conserving'') 
theories. \cite{BK61}
Higher-order approaches as the fluctuation-exchange approximation 
\cite{BSW89} are mainly applied to discrete lattice models while
for continuum systems, e.g.\ for the inhomogeneous electron gas, 
one has to be 
content with lowest-order theories as the GW method. \cite{SMH80,SMH82,AG98}

A second type of general methods is based on 
density-functional (DF) approaches. \cite{HK64,KS65}
Normally these aim at the inhomogeneous electron gas but can also 
be applied to Hubbard-type lattice models. \cite{SGN95}
Compared with the Green's-function formalism, there is a conceptually 
similar situation for DF approaches:
In the latter the ground-state energy $E$ (or the grand potential 
$\Omega$) \cite{Mer65}
is given as a functional of the (static) density ${\bf n}$.
The variational principle associated with the functional $E[{\bf n}]$ is 
rigorous but cannot be evaluated as $E[{\bf n}]$ is generally unknown.
In the standard local-density approximation (LDA) the (unknown
part of the) functional is replaced by an explicitly known but approximate 
functional which is taken from the homogeneous system.
For systems with weakly varying density the LDA should be justified.
Information on excitation properties is contained in dynamic response 
functions which are in principle accessible via time-dependent DF theory 
\cite{RG84} where the action $A$ is considered as a functional of the 
time-dependent density ${\bf n}$.
Again, the exact but unknown functional $A[{\bf n}]$ is approximated to
make it explicit and the variational principle is exploited afterwards.

The method proposed here rests on a variational principle which uses 
the electron self-energy ${\bf \Sigma}$ as the basic dynamic variable.
A new functional $\Omega[{\bf \Sigma}]$ is constructed which can be shown
to be stationary at the physical self-energy.
The main result is that the variational principle can be exploited without
any approximation of the functional dependence. 
Namely, a {\em rigorous} evaluation of the functional $\Omega[{\bf \Sigma}]$ 
is possible on a certain subspace of trial self-energies.
Trial self-energies must be representable within an exactly solvable
reference system sharing the same interaction with the original system.

This result has important consequences as it opens a route for constructing
a novel class of approximations.
Although the self-energy essentially contains the same information as the 
Green's function or the (time-dependent) density,
the new approach is conceptually contrary to the Green's-function approach 
and to the DF approach
as there is no approximation to be tolerated for the central functional.
Instead of approximating the functional itself, it is considered 
on a restricted domain.
The self-energy-functional approach is completely general and yields 
approximations which are non-perturbative, thermodynamically consistent 
and systematic. 
Opposed to numerical techniques directly applied to systems of finite
size, the self-energy-functional approach provides a variational or
self-consistent embedding of finite systems and thus yields results 
in the thermodynamical limit.
Such techniques are needed to construct phase diagrams from standard 
correlated lattice models.
A potentially fruitful field of application are systems with competing
types of order resulting from spin, charge or orbital correlations
as it is typical e.g.\ for numerous transition-metal oxides.
\cite{OM00,Geb97,BDN01}

In the present paper the approach is introduced and a number of general 
aspects are discussed in detail (Sec.\ \ref{sec:sft}).
To demonstrate its usefulness, two applications will be considered for 
the single-band Hubbard model:
In Sec.\ \ref{sec:dmft} it is shown that the dynamical mean-field 
theory (DMFT) \cite{GKKR96} can be recovered within the 
self-energy-functional approach, namely by choosing a decoupled set of 
impurity Anderson models as a reference system.
The DMFT generally requires the treatment of a quantum-impurity problem
with an infinite number of degrees of freedom ($n_{\rm s} = \infty$).
In Sec.\ \ref{sec:ed} a new approximation is discussed which is based 
on an impurity model with a {\em finite} number of degrees of freedom 
only and which approaches the DMFT for $n_{\rm s} = \infty$.
The method is closely related to the exact-diagonalization approach (ED).
\cite{CK94,SRKR94}
Opposed to the ED, however, thermodynamical consistency is guaranteed at
any stage of the approximation.
New approaches beyond the mean-field level will be discussed elsewhere.
The conclusions and an outlook are given in Sec.\ \ref{sec:con}.

\section{Self-energy-functional approach}
\label{sec:sft}

Consider a general Hamiltonian $H = H_0({\bf t}) + H_1 ({\bf U})$ with 
one-particle (``hopping'') parameters ${\bf t}$ and two-particle 
interaction parameters ${\bf U}$:
\begin{equation}
  H = \sum_{\alpha\beta} t_{\alpha\beta} 
  c_{\alpha}^\dagger c_{\beta}
  + \frac{1}{2} 
  \sum_{\alpha\beta\gamma\delta} U_{\alpha\beta\delta\gamma}
  c_{\alpha}^\dagger c^\dagger_{\beta} c_{\gamma} c_{\delta} \: .
\end{equation}
Here $\alpha,\beta,...$ refer to an orthonormal and complete set of 
one-particle basis states. 
We are interested in the equilibrium thermodynamics and in elementary 
one-particle excitations of the system for temperature $T$ and chemical 
potential $\mu$. 
This is described by the one-particle Green's function 
$G_{\alpha\beta}(i\omega)=\langle \langle c_\alpha ; c_\beta^\dagger 
\rangle \rangle$ of the imaginary fermionic Matsubara frequencies 
$i\omega = i(2n+1) \pi T$ with integer $n$. \cite{AGD64}
The Green's function can be calculated from the self-energy 
$\Sigma_{\alpha\beta}(i\omega)$ via the Dyson equation. 
Using a matrix notation, this reads as 
${\bf G} = {\bf G}_0 + {\bf G}_0 {\bf \Sigma} {\bf G}$
where ${\bf G}_0 = 1/(i\omega + \mu - {\bf t})$ is the ``free'' 
Green's function.
The self-energy is given by ${\bf \Sigma} = {\bf \Sigma}[{\bf G}] = 
T^{-1} \delta \Phi[\bf G] / \delta {\bf G}$,
where $\Phi[{\bf G}]$ is the so-called Luttinger-Ward functional.
\cite{LW60,BK61}
This allows to derive the Green's function from a variational principle:
One has $\delta {\Omega} [{\bf G}] / \delta {\bf G} = 0$ where
${\Omega} [{\bf G}] = \Phi[{\bf G}] + \mbox{Tr} \ln (- {\bf G}) - 
\mbox{Tr} (({\bf G}_0^{-1} - {\bf G}^{-1}) {\bf G})$ and using the
notation
$\mbox{Tr} \, {\bf A} = T \sum_{\omega,\alpha} A_{\alpha\alpha}(i\omega)$.
In general, however, the functional $\Phi[{\bf G}]$ is not known 
explicitly which prevents an evaluation of ${\Omega} [{\bf G}]$
for a given ${\bf G}$.
So-called conserving approximations \cite{BK61} provide an explicit but 
approximate functional $\Phi_{\rm pert.}[{\bf G}] \approx \Phi[{\bf G}]$.
However, these are weak-coupling approaches where a certain subclass of 
$\Phi$ diagrams is summed up.

Here a different but still rigorous variational principle is proposed
which is based on a functional ${\bf G} = {\bf G}[{\bf \Sigma}]$ defined 
as the inverse of ${\bf \Sigma} = {\bf \Sigma}[{\bf G}]$.
We can assume the latter to be invertible (locally) provided that the 
system is not at a critical point for a phase transition 
(see Appendix \ref{sec:gofs}).
Consider then:
\begin{equation}
  \Omega_{\bf t}[{\bf \Sigma}] \equiv 
  {\rm Tr} \ln (- ({\bf G}_0^{-1} - {\bf \Sigma})^{-1}) + 
  F[{\bf \Sigma}]
\label{eq:var}
\end{equation}
where $F[{\bf \Sigma}] \equiv \Phi[{\bf G}[{\bf \Sigma}]] - 
\mbox{Tr} ({\bf \Sigma} \, {\bf G}[{\bf \Sigma}])$ 
is the Legendre transform of $\Phi[{\bf G}]$.
The subscript ${\bf t}$ indicates the explicit ${\bf t}$ dependence
of $\Omega$ due to the free Green's function ${\bf G}_0$.
Using $T^{-1} \delta F[{\bf \Sigma}] / \delta {\bf \Sigma} = 
{\bf G}[{\bf \Sigma}]$, one finds that
\begin{equation}
  \delta \Omega_{\bf t} [{\bf \Sigma}] / \delta {\bf \Sigma} = 0 
  \; \Leftrightarrow \; {\bf G}[{\bf \Sigma}] 
  = ({\bf G}_0^{-1} - {\bf \Sigma})^{-1} \: .
\label{eq:stat}  
\end{equation}
Thus $\Omega_{\bf t}[{\bf \Sigma}]$ is stationary at the exact (physical) 
self-energy and its value is the exact grand potential of the system.
Again, the problem is that the functional $\Omega_{\bf t}[{\bf \Sigma}]$ 
is in general not known explicitly.

As the domain of the self-energy functional $\Omega_{\bf t}[{\bf \Sigma}]$ 
we define the class of all ${\bf t}'$ representable self-energies.
${\bf \Sigma}$ is termed ${\bf t}'$ representable, if there is a 
set of hopping parameters ${\bf t}'$ such that ${\bf \Sigma}$ is 
the exact self-energy of the model $H_0({\bf t}') + H_1 ({\bf U})$.
This implies that any self-energy in the domain of 
$\Omega_{\bf t}[{\bf \Sigma}]$ can be parameterized as 
${\bf \Sigma} = {\bf \Sigma}({\bf t}')$.
The interaction parameters ${\bf U}$ are taken to be fixed.
Suppose we are interested in the model $H = H_0({\bf t}) + H_1 ({\bf U})$.
Then the function 
$\Omega_{\bf t}({\bf t}') \equiv \Omega_{\bf t}[{\bf \Sigma}({\bf t}')]$ 
is stationary at ${\bf t}' = {\bf t}$. 
Thus $\partial \Omega_{\bf t}({\bf t}') / \partial {\bf t}' = 0$.

It is important to note that $F[{\bf \Sigma}]$ is universal:
The functional dependence is the same for any ${\bf t}$, i.e.\ it remains
unchanged for an arbitrary reference system $H'$ with the same 
interaction but modified hopping parameters:
$H' = H_0({\bf t}') + H_1 ({\bf U})$.
$F[{\bf \Sigma}]$ is universal as it is the Legendre tranform 
of $\Phi[{\bf G}]$ which in turn is universal because it can be 
constructed formally as the sum of all closed, irreducible, and 
renormalized skeleton diagrams which, apart from ${\bf G}$, include
the vertices ${\bf U}$ only.
Consequently, one has:
\begin{equation}
  \Omega_{{\bf t}'}[{\bf \Sigma}] = 
  {\rm Tr} \ln (- ({{\bf G}'_0}^{-1} - {\bf \Sigma})^{-1}) + 
  F[{\bf \Sigma}] \: ,
\label{eq:varp}
\end{equation}
for the reference system $H'$ with 
${{\bf G}'_0}^{-1} = i\omega + \mu - {\bf t}'$.
Combining Eqs.\ (\ref{eq:var}) and (\ref{eq:varp}), $F[{\bf \Sigma}]$ 
can be eliminated:
\begin{eqnarray}
  \Omega_{\bf t}[{\bf \Sigma}] 
  = \Omega_{{\bf t}'}[{\bf \Sigma}]
  &+& {\rm Tr} \ln (- ({\bf G}_0^{-1} - {\bf \Sigma})^{-1})
\nonumber \\   
  &-& {\rm Tr} \ln (- ({{\bf G}'_0}^{-1} - {\bf \Sigma})^{-1}) 
  \: .
\label{eq:felim}
\end{eqnarray}
Evaluating the functional $\Omega_{\bf t}[{\bf \Sigma}]$ for 
self-energies parameterized as ${\bf \Sigma} = {\bf \Sigma}({\bf t}')$,
one obtains:
\begin{equation}
   \Omega_{\bf t}[{\bf \Sigma}({\bf t}')] = \Omega' 
   + {\rm Tr} \ln (- ({\bf G}_0^{-1} - {\bf \Sigma}({\bf t}'))^{-1})
   - {\rm Tr} \ln (-{\bf G}') \: .
\label{eq:vvv}
\end{equation}
Here it has been used that 
$\Omega_{{\bf t}'}[{\bf \Sigma}({\bf t}')] = \Omega'$,
the exact grand potential of the reference system $H'$, and
$({{\bf G}'_0}^{-1} - {\bf \Sigma}({\bf t}'))^{-1} = {\bf G}'$, the exact
Green's function of $H'$.
Suppose that the reference system $H'$ is much simpler than the original 
system $H$ so that it can be solved exactly for any ${\bf t}'$ belonging
to a certain subspace of the entire space of hopping parameters.
The resulting Eq.\ (\ref{eq:vvv}) is remarkable, as it shows that the 
functional $\Omega_{{\bf t}}[{\bf \Sigma}]$ can be evaluated rigorously
for trial self-energies ${\bf \Sigma}={\bf \Sigma}({\bf t}')$ taken from
the reference system $H'$.

This is the main result.
Contrary to previous approaches (e.g.\ conserving theories, LDA), 
there is no need to approximate the functional dependence in a 
fundamental variational principle.
Approximations are constructed by searching for a stationary point of 
$\Omega_{{\bf t}}[{\bf \Sigma}]$ on a {\em restricted} set of trial 
self-energies ${\bf \Sigma}({\bf t}')$. 

The stationary point is determined by the Euler equation:
$\partial \Omega_{\bf t}[{\bf \Sigma}({\bf t}')] / \partial {\bf t}'=0$.
Calculating the derivative,
\begin{equation}
   T \sum_{\omega} \sum_{\alpha\beta}
   \left( 
   \frac{1}{{\bf G}_0^{-1} - {\bf \Sigma}({\bf t}')} 
   -  {\bf G}' \right)_{\beta \alpha} 
   \frac{\partial \Sigma_{\alpha\beta}({\bf t}')}
        {\partial {{\bf t}'}}
   = 0  \: . 
\label{eq:euler}
\end{equation}
Note that the equation involves, apart from ${\bf G}_0$, quantities of the
reference system $H'$ only.
The linear response of the self-energy of $H'$ due to a change of the
hopping ${\bf t}'$ can be calculated along the lines of Ref.\ \onlinecite{BK61}. 
It turns out that $\partial {\bf \Sigma}({\bf t}') / \partial {\bf t}'$ 
is given by a two-particle Green's function of $H'$.
Since ${\bf G}' = {\bf G}[{\bf \Sigma}({\bf t}')]$, the exact self-energy
of the system $H$ is determined by the condition that the bracket in 
(\ref{eq:euler}) be zero. 
Hence, one can consider Eq.\ (\ref{eq:euler}) to be obtained from the
{\em exact} equation that determines the ``vector'' ${\bf \Sigma}$ in 
the self-energy space through {\em projection} onto the hypersurface of 
${\bf t}'$ representable trial self-energies ${\bf \Sigma}({\bf t}')$ 
by taking the scalar product with vectors 
$\partial {\bf \Sigma}({\bf t}') / \partial {\bf t}'$ tangential to 
the hypersurface.

An analysis of the second derivative
$\partial^2 \Omega_{\bf t} [{\bf \Sigma}({\bf t}')] / 
\partial t'_{\alpha\beta} \partial t'_{\gamma\delta}$
shows that a stationary point is not an extremum point in general.
This feature is shared with the time-dependent DF approach, \cite{RG84}
the Green's-function approach \cite{BK61} and also with a recently considered 
variant. \cite{CK01}
Only in the static DF theory there is a convex (density) functional.
\cite{HK64,KS65,Mer65}
Nevertheless, the proposed self-energy-functional approach is systematic:
For any sequence of reference systems $H'$ including more and more 
degrees of freedom and converging to the original system $H$ there 
is, from the variational principle, a corresponding sequence of grand 
potentials 
which must converge to the exact
$\Omega = \Omega_{\bf t} [{\bf \Sigma}({\bf t})]$ 
as the subspace of trial self-energies increases 
and eventually includes the exact self-energy
${\bf \Sigma}({\bf t})$.

\section{Relation to the DMFT}
\label{sec:dmft}

Given an original model $H$, what could a suitable reference system 
$H'$ look like?
Consider, for example, $H$ to be the Hubbard model \cite{hub}
which is shown in Fig.~1a schematically: 
A filled dot represents a correlated site $i$ with on-site Hubbard 
interaction $U$, and a line connecting two sites $i$ and $j$ represents 
the nearest-neighbor hopping $t_{i,j}$.
The number of sites is $L \mapsto \infty$.
Fig.~1c shows a conceivable reference system $H'$. 
$H'$ is obtained from $H$ (Fig.~1a) by (i) adding to each correlated site 
$i$ a number of $n_{\rm s}-1$ uncorrelated (``bath'') sites 
$k=2,...,n_{\rm s}$ (open dots) which are disconnected from the
rest of the system,
by (ii) switching off the hopping $t_{i,j}$ 
between the correlated sites and (iii) switching on a hopping $V_{i,k}$ 
to the bath sites.
After step (i) the Hamiltonian Fig.~1b (in the figure $n_{\rm s}=5$)
has an enlarged Hilbert space but the same self-energy.
It is important to note that steps (i) - (iii) leave the interaction part 
unchanged and thus preserve the functional dependence $F[{\bf \Sigma}]$.
Actually, the system $H'$ is a set of $L$ decoupled single-impurity 
Anderson models (SIAM) \cite{And61} with $n_{\rm s}$ sites each.
Compared to $H$, the problem posed by $H'$ is strongly simplified.
This is achieved at the cost of restricting the set of trial self-energies.
In particular, as the correlated sites are decoupled in $H'$, the trial
self-energies are local: $\Sigma_{ij}(i\omega,{\bf t}') \propto \delta_{ij}$.
One has to consider $H'$ for arbitrary one-particle parameters, 
namely the on-site energies of the correlated (``c'') and of the bath sites
(``a''), $\epsilon^{(c)}_{i}$ and $\epsilon^{(a)}_{i,k}$, respectively,
and the hopping (``hybridization'') $V_{i,k}$ between them
and take these as variational parameters in the principle
$\delta \Omega_{\bf t}[{\bf \Sigma}({\bf t}')] = 0$.

\begin{figure}[t]
\centerline{\includegraphics[width=90mm]{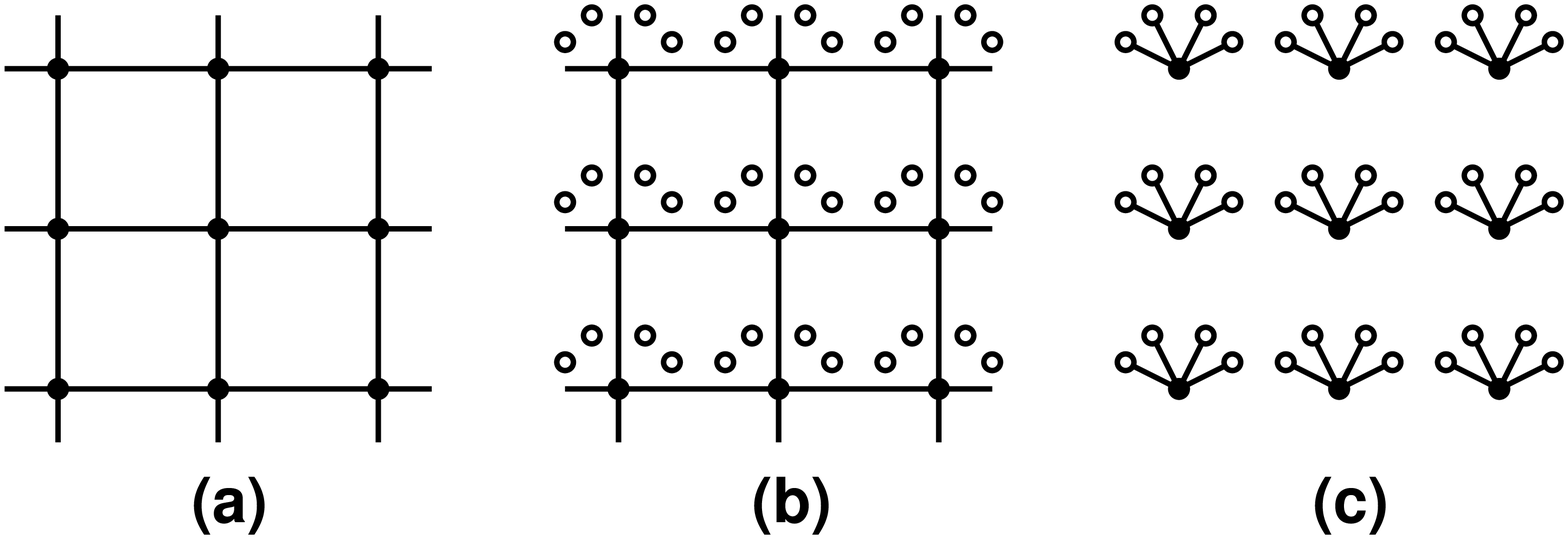}}
\caption{
Schematic representation of the Hubbard model $H$ (a),
an equivalent model (b), 
and a possible reference system $H'$ (c).
See text for discussion.
}
\label{fig:hhp}
\end{figure}

Let us discuss the case $n_{\rm s} \mapsto \infty$.
For a homogeneous phase of the (translationally invariant) original system, 
$\Omega_{\bf t}[{\bf \Sigma}({\bf t}')]$ will be stationary at a homogeneous 
set of variational parameters:
${\bf t}'=\{ \epsilon^{(c)}_{i}, \epsilon^{(a)}_{i,k}, V_{i,k} \}=
\{ \epsilon^{(c)}, \epsilon^{(a)}_{k}, V_{k}\}$.
Consequently, it is sufficient to consider one SIAM only.
As the different equivalent SIAM's are spatially decoupled, not only the
self-energy but also its linear response is local:
$\partial \Sigma_{ij}({\bf t}') / \partial {\bf t}' \propto \delta_{ij}$.
To solve the Euler equation (\ref{eq:euler}), it is thus sufficient to 
fulfill the ``locally projected'' equation
\begin{equation}
   \left( \frac{1}{{\bf G}_0^{-1}(i\omega) 
   - {\bf \Sigma}(i\omega)} \right)_{ii}
   = G'_{ii}(i\omega) \: .
\label{eq:dmft}
\end{equation}
This is just the self-consistency equation of the DMFT: \cite{GKKR96}
The SIAM parameters have to be found such that the on-site (``impurity'')  
Green's function at a correlated site $i$ coincides with
the on-site Green's function of the Hubbard model which is calculated
from ${\bf G}_0$ and the (``impurity'') self-energy of $H'$ 
by means of the Dyson equation.
Therefore, one can state that the DMFT 
(as an approximation for any finite-dimensional 
system or as the exact theory in infinite dimensions) is recovered as a 
stationary point of $\Omega_{\bf t}[{\bf \Sigma}]$ when restricting the 
search to local self-energies representable by a SIAM.

Within the DMFT the computation of the self-energy requires an
iterative procedure: ${\bf \Sigma}_{\rm old} \mapsto {\bf \Sigma}_{\rm new}$.
Here it turns out that this corresponds to a certain (discrete) path
on the hypersurface of SIAM trial self-energies.
Convergence of the iteration 
(${\bf \Sigma}_{\rm old} = {\bf \Sigma}_{\rm new}$),
however, is by no means guaranteed physically but depends on the 
contracting properties of the map
${\bf \Sigma}_{\rm old} \mapsto {\bf \Sigma}_{\rm new}$.
The self-energy-functional approach offers an alternative as instead of
solving Eq.\ (\ref{eq:dmft}) one may calculate 
$\Omega_{\bf t}[{\bf \Sigma}({\bf t}')]$ by Eq.\ (\ref{eq:vvv}) and 
determine the stationary point.
Hence, the DMFT can also be obtained by a more {\em direct} computation
avoiding any iterations -- similar (in this respect) to the 
random-dispersion approximation. \cite{NG99}
Note that in case of more than a single stationary point there is also an 
equally direct access to metastable phases.

For any inhomogeneous situation, Eq.\ (\ref{eq:dmft}) represents a system 
of self-consistency equations to fix the parameters of non-equivalent 
impurity models labeled by the site index $i$.
The models can be solved independently but are coupled indirectly due to 
the matrix inversion in (\ref{eq:dmft}). 
This exactly recovers the DMFT generalized to systems with reduced 
translational symmetry. \cite{PN99c,PN99d}

\section{A consistent ED method}
\label{sec:ed}

A brief discussion of two limiting cases of the Hubbard model may be 
instructive. Consider the band limit with $U=0$ first. 
Here $H = \sum_{ij\sigma} t_{ij} c_{i\sigma}^\dagger c_{j\sigma}$ describes 
a system of non-interacting electrons.
This case is exceptional as obviously the functional $F[{\bf \Sigma}] 
\equiv 0$ and therefore $\Omega_{\bf t}[{\bf \Sigma}] = {\rm Tr} \ln 
(- ({\bf G}_0^{-1} - {\bf \Sigma})^{-1})$. 
Any valid reference system $H'$ must have the same (i.e.\ a vanishing) 
interaction part as $H$, and thus ${\bf \Sigma}({\bf t}') \equiv 0$ and 
$\Omega_{\bf t}[{\bf \Sigma}({\bf t}')] = {\rm Tr} \ln (-{\bf G}_0^{-1})$, 
the exact grand potential for non-interacting electrons.

The atomic limit, $H = \sum_{i\sigma} ( t_{0} c_{i\sigma}^\dagger c_{i\sigma}
+ (U/2) n_{i\sigma} n_{i-\sigma})$ is more interesting as $\Phi[{\bf G}]$
and $F[{\bf \Sigma}]$ cannot be constructed explicitly.
Within the self-energy-functional approach one has to compute
${\bf \Sigma({\bf t}')}$, ${\bf G}'$, and $\Omega'$ for a suitable reference
system $H'$ and to insert into Eq.\ (\ref{eq:vvv}) for optimization.
The only meaningful choice for the reference system is $H'=H$ in this case.
Obviously, this yields the exact solution.
Generally, whenever the original model $H$ is exactly solvable, the 
choice $H'=H$ will do.

For a non-trivial model $H$, the choice $H'=H$ is useless for a practical 
computation.
Any simplified reference system, however, yields a consistent approximation.
The case of the DMFT discussed in Sec.\ \ref{sec:dmft} is an illustrative
example.
On the other hand, in the context of the DMFT actually both, $H$ and $H'$, 
are highly non-trivial models, and further approximations or large-scale 
numerics are needed to treat the reference system $H'$.
More simple approximations for the Hubbard model which are still constistent
are generated by considering reference systems with a {\em finite} number of 
degrees of freedom. 
The reference system of Fig.~1c with $n_{\rm s} < \infty$ is an interesting
example which shall be discussed in the following.
For small $n_{\rm s}$ one can easily obtain numerical results as a complete 
diagonalization of $H'$ is feasible.

\begin{figure}[t]
\centerline{\includegraphics[width=65mm]{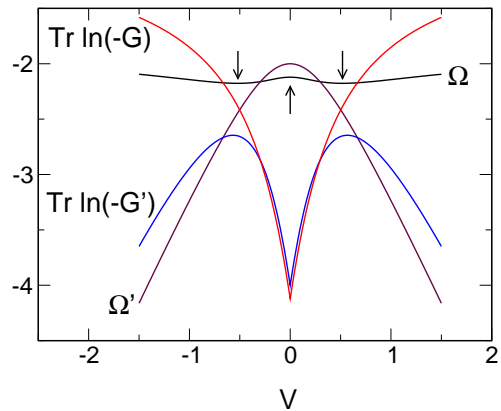}}
\caption{
Grand potential $\Omega$ (per lattice site) and the different contributions
(per lattice site) according to Eq.\ (\ref{eq:vvv}): $\Omega'$, 
${\rm Tr} \ln (- ({\bf G}_0^{-1} - {\bf \Sigma}({\bf t}'))^{-1})$, 
and ${\rm Tr} \ln (-{\bf G}')$ for $U=4$, $T=0$, and $\mu=U/2$ (half-filling)
as functions of $V$.
Stationary points (arrows) at $V=\pm 0.519$ (metal) and at $V=0$ (insulator).
$\epsilon_c=0$ and $\epsilon_a=2$.
}
\label{fig:e1}
\end{figure}

Consider the Hubbard model 
\begin{equation}
  H = \sum_{ij\sigma} t_{ij} c_{i\sigma}^\dagger c_{j\sigma} 
  + \frac{U}{2} \sum_{i\sigma} n_{i\sigma} n_{i-\sigma}
\end{equation}
at temperature $T=0$ and chemical potential $\mu = U/2$.
For symmetric conditions this implies half-filling.
The Hamiltonian of the reference system is given by $H'= \sum_i H'(i)$ with
\begin{eqnarray}
  H'(i) &=& \sum_{\sigma} \epsilon^{(c)}_{i} c_{i\sigma}^\dagger c_{i\sigma} 
            + \frac{U}{2} \sum_{\sigma} n_{i\sigma} n_{i-\sigma} 
	    \nonumber \\
	&+& \sum_{\sigma,k=2}^{n_{\rm s}} \epsilon^{(a)}_{i,k} a^\dagger_{ik\sigma} a_{ik\sigma}
	 +  \sum_{\sigma,k} \left( V_{i,k} c_{i\sigma}^\dagger a_{ik\sigma}  + \mbox{h.c.} \right) \: .
	 \nonumber \\
\label{eq:siam2}
\end{eqnarray}
For the sake of simplicity we consider a homogeneous paramagnetic phase and
the most simple case $n_{\rm s}=2$ where one is left with three independent 
variational parameters only, namely $V \equiv V_{i,k=2}$,
$\epsilon_a \equiv \epsilon^{(a)}_{i,k=2}$, and 
$\epsilon_c \equiv \epsilon^{(c)}_{i}$.
The computation of the different contributions to the grand potential,
Eq.\ (\ref{eq:vvv}), is straightforward:
Diagonalization of $H'$ yields the ground-state energy $E_0'$ and
$\Omega' = E_0' -\mu \langle N' \rangle$ as well as the excitation energies,
the ground state and the excited states.
The Green's function ${\bf G}'$ and the free Green's function 
${\bf G}'_0$ can be computed from their respective Lehmann representations.
The self-energy of the reference system is obtained as 
${\bf \Sigma}({\bf t}') = {{\bf G}'_0}^{-1} - {{\bf G}'}^{-1}$.
Since the self-energy is local, as in the DMFT, the lattice structure 
enters via the free ($U=0$) density of states only.
Therefore, the ${\bf k}$-sum which appears in the first trace in 
Eq.\ (\ref{eq:vvv}) can be performed conveniently by a one-dimensional 
density-of-states integration.
A semi-elliptical free density of states with the band width $W=4$ is
used for the calculations. 
This sets the energy scale for the results discussed below.

Fig.\ \ref{fig:e1} shows the grand potential $\Omega$ and the three different
contributions as functions of $V$. 
The interaction is kept fixed at $U=W=4$.
The remaining variational parameters are set to $\epsilon_c=0$ and 
$\epsilon_a=U/2=2$, as required by particle-hole symmetry.
Each of the three contributions strongly depends on $V$ and none of them
has a stationary point at a finite $V\ne 0$. 
Two of them show a singular behavior at $V=0$.
Contrary, the resulting $\Omega$ is regular for any $V$ and shows a much 
weaker $V$ dependence.
There are three stationary points which are indicated by the arrows.
The maximum at $V=0$ corresponds to an insulating phase since 
${\Sigma}(\omega)$ for $n_{\rm s}=2$ and $V=0$ is the Hubbard-I self-energy 
which implies a vanishing spectral density 
$-(1/\pi)\mbox{Im}{\bf G}(\omega+i0^+)$ at $\omega=0$.
The minima at $V=\pm 0.519$ correspond to a metallic phase.
$\Omega$ as well as the different contributions are symmetric functions
of $V$.
As ${\Sigma}(V,\omega)={\Sigma}(-V,\omega)$, however, this symmetry is trivial
and does not yield an additional physical phase (see also Appendix 
\ref{sec:gofs}).
Due to the lower $\Omega$ the metallic phase is stable as compared to the
insulating one.

\begin{figure}[t]
\centerline{\includegraphics[width=78mm]{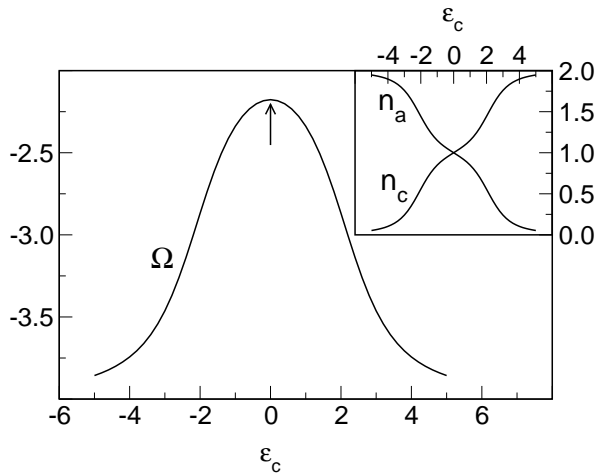}}
\caption{
$\Omega$ as a function of $\epsilon_{\rm c}$ for $V=0.519$ (metal) and 
$\epsilon_{\rm a}=2$. $U=4$.
Inset: impurity- and bath-orbital filling, $n_{\rm c}$ and
$n_{\rm a}$, as functions of $\epsilon_{\rm c}$. 
}
\label{fig:e2}
\end{figure}

\begin{figure}[t]
\centerline{\includegraphics[width=78mm]{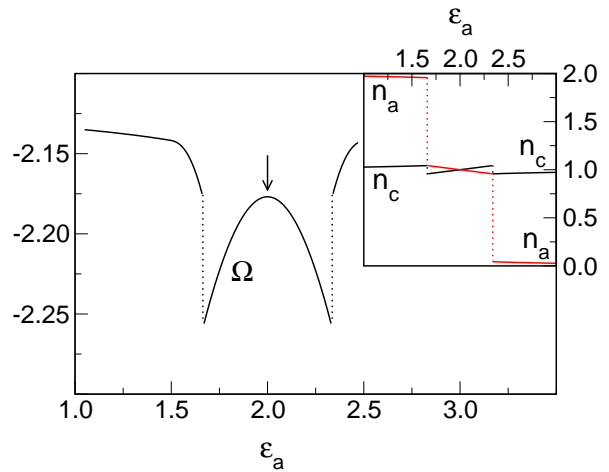}}
\caption{
$\Omega$ and the impurity- and bath-orbital filling, $n_{\rm c}$ and
$n_{\rm a}$, as functions of $\epsilon_{\rm a}$ for
$V=0.519$ and $\epsilon_{\rm c}=0$. 
$U=4$.
}
\label{fig:e3}
\end{figure}

The minimum at $V=0.519$ is actually a saddle point if the entire space 
of variational parameters is considered.
This is demonstrated by Fig.\ \ref{fig:e2} which shows $\Omega$ as a function
of $\epsilon_c$ for fixed $V=0.519$ and $\epsilon_a=2$.
While $\Omega(V)$ is at a minimum for $V=0.519$, $\Omega(\epsilon_c)$ is at 
a {\em maximum} for $\epsilon_c=0$. 
In the $(V,\epsilon_c)$ space one therefore encounters a saddle point.
As already noted in Sec.\ \ref{sec:sft}, there is no reason to expect an
extremum in general.
It is worth mentioning that stationarity at $\epsilon_c = 0$ is consistent
with the requirements of particle-hole symmetry.
For any $\epsilon_c \ne 0$ the impurity model is asymmetric.
This can be seen from the inset where the average occupations of the impurity
and of the bath site are plotted as functions of $\epsilon_c$.
The total particle number $\langle N' \rangle = \sum_\sigma 
(\langle c_{\sigma}^\dagger c_{\sigma} \rangle + 
\langle a^\dagger_{\sigma} a_{\sigma}\rangle) = n_{\rm c} + n_{\rm a}$
($i$ and $k=2$ fixed) is constant: $\langle N' \rangle = 2$.

With respect to the third variational parameter $\epsilon_a$, the grand 
potential $\Omega$ is at a maximum for $\epsilon_a = 2 = \mu$, see Fig.\ 
\ref{fig:e3}.
Again, this value is required by particle-hole symmetry.
If $\epsilon_a$ exceeds a certain critical value (away from the stationary 
point), the ground-state of the reference system $H'$ no longer lies within
$N'=2$ subspace but is found in the $N'=1$ or $N'=3$ subspace, respectively
(see inset of Fig.\ \ref{fig:e3}).
While $\Omega'$ is continuous at the level crossing, the symmetry of the 
ground state changes.
Consequently, there is a discontinuous change of the trial self-energy
which implies a discontinuous change of $\Omega$.

Consider now the {\em original} model at slightly modified parameters,
e.g.\ $U$, $\mu$, or $T$.
Clearly, the stationary point of $\Omega$ will be expected then at 
slightly different values of the variational parameters $V$, $\epsilon_c$, 
$\epsilon_a$.
This implies that all physical quantities which derive from the 
thermodynamical potential $\Omega$ will be continuous functions of 
the (original) model parameters in general -- irrespective of the fact 
that the reference system includes a finite number of degrees of freedom
only:
It is a typical feature of any mean-field approach that results are directly
provided for the thermodynamical limit.
A discontinuous jump of $\Omega$ due to a symmetry change of the ground state
of the reference system (see Fig.\ \ref{fig:e3}) usually occurs away from 
stationarity and is thus irrelevant physically.
It is conceivable, however, that the stationary point moves to a point of 
discontinuity as a function a parameter of the original model.
In this case the approach would generate an artifact which is a 
reminiscence of the finiteness of $H'$. 

\begin{figure}[t]
\centerline{\includegraphics[width=80mm]{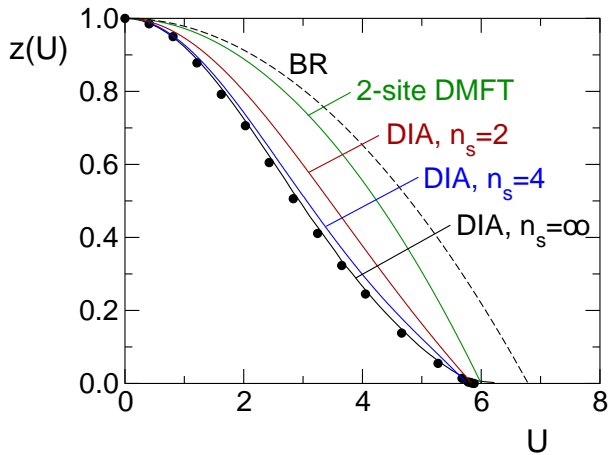}}
\caption{
$U$ dependence of the quasi-particle weight 
$z = 1 / (1-d \Sigma(\omega = 0)/d\omega)$ within different approximations.
BR: Brinkman-Rice (Gutzwiller) approach. \cite{BR70}
2-site DMFT: a non-variational two-site approach. \cite{Pot01}
DIA, $n_{\rm s}=2$: self-energy-functional approach with a reference 
system $H'$ consisting of decoupled two-site impurity models.
DIA, $n_{\rm s}=4$: dynamical-impurity approximation with $n_{\rm s}=4$ 
sites.
DIA, $n_{\rm s}=\infty$: DMFT limit (circles: NRG, \cite{Bul99} line: ED
using 8 sites).
}
\label{fig:zofu}
\end{figure}

So far we discussed the case $U=4=W$ only.
As a function of $U$ the half-filled paramagnetic Hubbard model at $T=0$
is expected to undergo a transition from a metal to a Mott-Hubbard insulator.
\cite{Geb97}
This is marked by a divergence of the effective mass or, equivalently, by a 
vanishing quasi-particle weight $z=1/(1-d \Sigma(\omega)/d\omega|_{\omega=0})$
as $U$ approaches a critical value $U_{\rm c}$ from below. 
\cite{GKKR96,BR70}
The result for $z(U)$ as obtained by the use of the two-site reference system 
(``dynamical impurity approximation'', DIA with $n_{\rm s}=2$) is shown in 
Fig.\ \ref{fig:zofu}.
As there are less degrees of freedom contained in $H'$, the approximation 
should be considered to be {\em inferior} as compared to the full DMFT the
results of which are in Fig.\ \ref{fig:zofu}, too.
It is remarkable that the simple $n_{\rm s}=2$-DIA (which requires the 
diagonalization of a dimer model only) yields $z(U)$ in an almost quantitative
agreement with the full DMFT.

The results of the $n_{\rm s}=2$-DIA may also be compared with those of
the recently developed ``linearized'' or ``two-site'' DMFT \cite{BP00,Pot01}
where the Hubbard model is mapped onto the two-site SIAM (\ref{eq:siam2})
by means a strongly simplified self-consistency condition.
As compared to the two-site DMFT, the present self-energy-functional approach
not only represents a clear conceptual improvement but also improves the 
actual results for $z(U)$ and for $U_{\rm c}$ (see Fig.\ \ref{fig:zofu} 
and note that $U_{\rm c}=6$ within the linearized DMFT,
$U_{\rm c}=5.8450$ within the DIA for $n_{\rm s}=2$, and 
$U_{\rm c} = 5.84$ and $U_{\rm c} = 5.88$ from numerical evaluations 
\cite{MSK+95,Bul99} of the full DMFT).


The self-energy-functional approach with reference system of Fig.~1c and 
small $n_{\rm s}$ actually represents a new variant of the 
DMFT-exact-diagonalization method (ED). \cite{CK94,SRKR94}
As compared to previous formulations of the ED, the convergence with respect 
to $n_{\rm s}$ appears to be faster: 
Compare the results for $n_{\rm s}=2$, $n_{\rm s}=4$ and $n_{\rm s}=\infty$ 
(full DMFT) in Fig.\ \ref{fig:zofu} with those of Ref.\ \onlinecite{CK94}.
More important, however, there is no need for a fit procedure in the present
approach; any arbitrariness in the method to find the SIAM parameters is 
avoided completely.
Furthermore, consistent results will be obtained for any finite $n_{\rm s}$ 
while in the usual ED this can be expected in the DMFT limit 
$n_{\rm s} \mapsto \infty$ only.

\section{Conclusions and outlook}
\label{sec:con}

Concluding, the proposed self-energy-functional method is a systematic scheme 
for the construction of new non-perturbative and consistent approximations for 
extended systems of interacting fermions.
For Hubbard-type lattice models with on-site interaction several relations
to and generalizations of existing approaches are obtained immediately.
The numerical results obtained by considering a rather simple reference 
system clearly demonstrate the practicability of the theory.
Its generality promises that the approach may successfully be applied 
also in different contexts: 

For a Hubbard-type system including $M>1$ orbitals per site, a consistent
DMFT can only be defined when using $M$ baths.
There is no such necessity within the self-energy-functional approach.
While clearly the optimal local approximation requires $M$ baths,
any $M'<M$ will nevertheless lead to a fully consistent approximation.
This represents an interesting option for numerical studies of multi-band 
systems.
Non-local trial self-energies can be constructed by grouping the sites 
into identical clusters of finite size $N_{\rm s}$, switching off the 
inter-cluster hopping and treating the intra-cluster hopping as variational 
parameters.
Each of the $N_{\rm s}$ sites in a cluster can be coupled to $n_{\rm s}-1$ 
additional bath sites.
The relation of such an approach to cluster extensions of the DMFT
\cite{HTZ+98,KSPB01} and to the cluster perturbation theory 
\cite{GV93,SPPL00,SPP02}
will be the subject of a forthcoming paper.
Extensions and applications of the method to continuous models 
(inhomogeneous electron gas) and to Bose systems deserve 
further investigations.


\acknowledgments
The work is supported by the Deutsche Forschungsgemeinschaft 
(Sonderforschungsbereich 290).

\appendix

\section{The functional ${\bf G}[{\bf \Sigma}]$}
\label{sec:gofs}

For the definition of ${\bf G}[{\bf \Sigma}]$, invertibility of the 
functional ${\bf \Sigma}[{\bf G}]$ is required.
The {\em local} invertibility of ${\bf \Sigma}[{\bf G}]$ is controlled 
by the Jacobian $\Gamma_{\alpha\beta'\alpha'\beta}(i\omega,i\omega') =
\delta{\Sigma_{\alpha\beta}}(i\omega) / 
\delta{G_{\alpha'\beta'}}(i\omega')$.
The two-particle self-energy \cite{BK61} 
$\bf \Gamma = \delta{\bf \Sigma} / \delta {\bf G}$ can be assumed to 
be non-singular in general.

For a further analysis we need the following 
{\em lemma:} Consider the interaction ${\bf U}$ to be fixed. Then
two different Green's functions ${\bf G}_1$ and ${\bf G}_2$ must result from 
two different sets of one-particle parameters ${\bf t}'_1$ and ${\bf t}'_2$, 
respectively. The {\em proof} is straightforward:
Consider the high-frequency expansion of the Green's function
$G_{\alpha\beta}(\omega)=\sum_{n=1}^\infty M_{\alpha\beta}^{(n)} \omega^{-n}$.
The coefficients are given by the moments 
$M_{\alpha\beta}^{(n)} = \int d\omega \: \omega^n (-1/\pi) \, \mbox{Im} \, 
G_{\alpha\beta}(\omega+i0^+) = 
\langle [ {\cal L}^n c_\alpha , c_\beta ]_+\rangle$ with
${\cal L O} \equiv [{\cal O} , H]_-$. Using the symmetry
$U_{\alpha\beta\gamma\delta}=U_{\beta\alpha\delta\gamma}$ one has:
\begin{eqnarray}
  G_{\alpha\beta}(\omega) &=& \delta_{\alpha\beta} \frac{1}{\omega} 
\nonumber \\
  &+& \Big( t'_{\alpha\beta} +
  \sum_{\gamma\delta} \left( 
  U_{\alpha\gamma\beta\delta} - U_{\alpha\gamma\delta\beta} 
  \right) 
  \langle c^\dagger_\gamma c_\delta \rangle 
  \Big) 
  \frac{1}{\omega^2}
\nonumber \\
  &+& {\cal O}(\omega^{-3})
\end{eqnarray}
Now ${\bf t}'_1 \ne {\bf t}'_2$ implies the $\omega^{-2}$ coefficients
to be different because
\begin{equation}
  \langle c^\dagger_\alpha c_\beta \rangle 
  = - \frac{1}{\pi} \int_{-\infty}^\infty d\omega \: 
  \frac{1}{e^{\omega/T}+1}
  \, \mbox{Im} \, G_{\beta\alpha}(\omega+i0^+) \: .
\end{equation}
Consequently, we must have ${\bf G}_1 \ne {\bf G}_2$.

The lemma shows that the relation ${\bf t}' \leftrightarrow {\bf G}$ is 
one-to-one. Consequently, we can write 
${\bf \Sigma}[{\bf G}] = {\bf \Sigma}[{\bf t}'[{\bf G}]]$ and 
$\bf \Gamma = \delta{\bf \Sigma} / \delta {\bf G} =
\delta{\bf \Sigma} / \delta {\bf t}' \cdot 
\delta{\bf t}' / \delta {\bf G}$ with a non-singular Jacobian 
$\delta{\bf t}' / \delta {\bf G}$.
Hence, a singular $\bf \Gamma = \delta{\bf \Sigma} / \delta {\bf G}$
implies a singular $\delta{\bf \Sigma} / \delta {\bf t}'$ and vice versa.
However, $\delta{\bf \Sigma} / \delta {\bf t}'$ is just the ``projector'' 
in the Euler equation (\ref{eq:euler}). 
We conclude that local non-invertibility of the functional
${\bf \Sigma}[{\bf G}]$ at ${\bf G}={\bf G}({\bf t}')$ is indicated by 
$\partial {\bf \Sigma}[{\bf t}'] / \partial t'_{\bf n} = 0$ with
$t'_{\bf n} = {\bf t}' \cdot {\bf n}$ for a certain ``direction'' ${\bf n}$ 
in the space of hopping paramters.
For such a direction, the Euler equation (\ref{eq:euler}) would be satisfied
{\em trivially}.

Referring to the present numerical results, one can state that generally 
the projector $\delta{\bf \Sigma} / \delta {\bf t}'$ is non-singular in 
fact, as has been expected. 
There is one exception, however, namely points in the hopping-parameter 
space where one or more bath sites are {\em decoupled} from the rest of 
the system (vanishing hybridization $V$).
Here the one-particle energy of a decoupled bath site can be varied
without changing the trial self-energy.
Even for this case, however, there are no formal difficulties with the 
inverse functional ${\bf G}[{\bf \Sigma}]$:
To ensure the local invertibility of ${\bf \Sigma}[{\bf G}]$, one simply 
has to restrict the space of variational parameters ${\bf t}'$ by excluding
the one-particle energies of the decoupled bath sites, i.e.\ one has to focus
on the physically relevant parameters.
This implies a respective restriction of the space of ${\bf t}'$-representable
Green's functions ${\bf G}({\bf t}')$ and ensures the local invertibility of 
${\bf \Sigma}[{\bf G}]$ on the restricted domain.
Similarly, a restriction of the ${\bf t}'$ space becomes necessary to
ensure the local invertibility of ${\bf \Sigma}[{\bf G}]$ in case of
a system where the self-energy is trivial (as e.g.\ for a model of 
spinless fermions with nearest-neighbor Coulomb interaction in the limit 
of infinite spatial dimensions where the self-energy is given by the Hartree 
term, cf.\ Ref.\ \onlinecite{MH89b}).

Finally, it should be mentioned that generally the functional 
${\bf \Sigma}[{\bf G}]$ cannot be inverted {\em globally}.
Consider, for example, the Hubbard model on the infinite-dimensional 
hypercubic lattice with nearest-neighbor hopping $t$ at half-filling. 
Due to manifest particle-hole symmetry, a sign change of the hopping 
$t \mapsto -t$ leaves the (local) self-energy invariant but transforms
(the non-local elements of) the Green's function ${\bf G}$.
We conclude that ${\bf G}[{\bf \Sigma}]$ in general cannot be defined 
uniquely.
Due to this non-uniqueness, and also due to non-linearity, there may be 
more than a single solution of Eq.\ (\ref{eq:stat}).
However, this does not cause any problem since 
for any $\bf \Sigma$ satisfying (\ref{eq:stat}) we have:
\begin{eqnarray}
  {\bf G}[{\bf \Sigma}] &=& ({\bf G}_0^{-1} - {\bf \Sigma})^{-1}
\nonumber \\  
  \Rightarrow \qquad
  {\bf \Sigma} &=& {\bf \Sigma}({\bf G}_0^{-1} - {\bf \Sigma})^{-1}
\nonumber \\  
  \Rightarrow \qquad
  {\bf \Sigma} &=& {\bf \Sigma}[{\bf G}] \quad \mbox{and} \quad
  {\bf G} = ({\bf G}_0^{-1} - {\bf \Sigma})^{-1} \: .
\end{eqnarray}
This means that ${\bf \Sigma}$ is given by the (formal) sum of all 
skeleton diagrams built up by a propagator ${\bf G}$ which is constructed 
via the Dyson equation from the same ${\bf \Sigma}$ in turn.
Hence, any stationary point should be regarded as a physically meaningful 
solution. Among different physical solutions (corresponding e.g.\ to 
different phases) the minimum grand potential selects the stable one.

\end{document}